\documentclass[conference]{IEEEtran}
\IEEEoverridecommandlockouts
\usepackage{cite}
\usepackage{amsmath,amssymb,amsfonts}
\usepackage{algorithmic}
\usepackage{graphicx}

\usepackage[T1]{fontenc}
\usepackage[utf8]{inputenc}
\usepackage{authblk}

\usepackage{orcidlink}

\usepackage{textcomp}
\usepackage{xcolor}
\usepackage[font=footnotesize,labelfont=bf]{caption}

\usepackage{tabularx}
\usepackage{dblfloatfix} % get table @bottom of page
\usepackage{booktabs}
\usepackage{paralist}
\usepackage{enumitem}
\usepackage[export]{adjustbox}
\usepackage{multicol}
\usepackage{gensymb}
\usepackage{steinmetz}
\usepackage{mathtools}
\usepackage{siunitx}
\usepackage{wrapfig}
\usepackage{subfig}
\usepackage[english]{babel}
\usepackage{blindtext}
\usepackage{setspace}

\newcommand{\distance}{-1.5em}

\def\BibTeX{{\rm B\kern-.05em{\sc i\kern-.025em b}\kern-.08em
    T\kern-.1667em\lower.7ex\hbox{E}\kern-.125emX}}

\let\OLDthebibliography\thebibliography
\renewcommand\thebibliography[1]{
  \OLDthebibliography{#1}
  \setlength{\parskip}{0pt}
  \setlength{\itemsep}{0pt plus 0.3ex}
}

\hypersetup{hidelinks}

\usepackage[absolute]{textpos}

\newcommand{\copyrightbar}{%
  \setlength{\fboxrule}{.4pt}   % frame thickness
  \setlength{\fboxsep}{4pt}     % inner padding
  \fcolorbox{black}{gray!15}{%
    \parbox{\dimexpr\textwidth-2\fboxsep-2\fboxrule}{%
      \centering\scriptsize
      © 2025 IEEE. Personal use of this material is permitted. Permission from IEEE must be obtained
      for all other uses, in any current or future media, including reprinting/republishing this
      material for advertising or promotional purposes, creating new collective works, for resale or
      redistribution to servers or lists, or reuse of any copyrighted component of this work in other
      works.\\[3pt]%
      DOI: \href{https://doi.org/10.1109/ECCE55643.2024.10861318}{10.1109/ECCE55643.2024.10861318}%
    }%
  }%
}

\begin{document}

\captionsetup[figure]{labelfont={bf},name={Fig.},labelsep=colon}
\captionsetup[table]{labelfont={bf},name={TABLE},labelsep=colon}

\title{Optimization-Based Comparative System Evaluation of Single and Dual Traction Inverters with Focus on Partial Load Efficiency and Chip Area\\
}

% \author{\IEEEauthorblockN{1\textsuperscript{st} Christoph Sachs}
% \IEEEauthorblockA{University of Applied Sciences Esslingen \\
% Faculty of Mobility and Technology \\
% Esslingen, Germany \\
% Christoph.Sachs@hs-esslingen.de}
% \and
% \IEEEauthorblockN{2\textsuperscript{nd} Fabian Stamer}
% \IEEEauthorblockA{Bosch GmbH\\
% Renningen, Germany \\
% Fabian.Stamer@de.bosch.com}
% \and
% \IEEEauthorblockN{3\textsuperscript{rd} Jan Allgeier}
% \IEEEauthorblockA{Bosch GmbH\\
% Renningen, Germany \\
% Jan.Allgeier@de.bosch.com}

% \and
% \IEEEauthorblockN{4\textsuperscript{th} Duleepa Thrimawithana}
% \IEEEauthorblockA{The University of Auckland \\
% Dept. of Electrical, Computer and Software Engineering\\
% Auckland, New Zealand \\
% d.thrimawithana@auckland.ac.nz}
% \and
% \IEEEauthorblockN{5\textsuperscript{th} Martin Neuburger}
% \IEEEauthorblockA{University of Applied Sciences Esslingen \\
% Faculty of Mobility and Technology \\
% Esslingen, Germany \\
% Martin.Neuburger@hs-esslingen.de}
% }

\author[1,2,*]{Christoph Sachs} 
\author[3]{Fabian Stamer}
\author[3]{Jan Allgeier}
\author[4]{Duleepa Thrimawithana}
\author[1]{Martin Neuburger}
\affil[1]{\textit{Faculty of Mobility and Technology, Esslingen University, Germany}}
\affil[2]{\textit{Institute for System Dynamics, University of Stuttgart, Germany}}
\affil[3]{\textit{Robert Bosch GmbH, Germany}}
\affil[4]{\textit{Dept. of Electrical, Computer and Software Engineering, The University of Auckland, New Zealand}}
\affil[*]{christoph.sachs@hs-esslingen.de}

% \author{
%     Christoph Sachs\IEEEauthorrefmark{1}\IEEEauthorrefmark{2}\,\orcidlink{0009-0005-5559-9342}, 
%     Fabian Stamer\IEEEauthorrefmark{3}\,\orcidlink{0000-0003-4663-1919}, 
%     Jan Allgeier\IEEEauthorrefmark{3}, 
%     Duleepa J. Thrimawithana\IEEEauthorrefmark{4}\,\orcidlink{0000-0002-4291-5220}

%     and 
%     Martin Neuburger\IEEEauthorrefmark{1}\,\orcidlink{0000-0002-3867-011X}
%     \\ 
%     \\
%     %\small
%     \IEEEauthorrefmark{1}\textit{Faculty of Mobility and Technology, Esslingen University, Germany,}\\
%     \IEEEauthorrefmark{2}\textit{Institute for System Dynamics, University of Stuttgart, Germany,}\\
%     \IEEEauthorrefmark{3}\textit{Robert Bosch GmbH, Germany,} \\
%     \IEEEauthorrefmark{4}\textit{Dep. of Electrical, Computer and Software Engineering, The University of Auckland, New Zealand}
% }

\maketitle

% % --- place right after \title/\author (before \maketitle for conferences) ----
% \IEEEpubid{\begin{minipage}[b]{\textwidth}\copyrightbar\end{minipage}}
% \IEEEpubidadjcol     % fills the gap LaTeX would otherwise leave

\begin{textblock*}{\textwidth}(40pt,730pt)  % Adjust 250pt as needed
\copyrightbar
\end{textblock*}

\begin{abstract}
The transition to electric transportation demands efficient and cost-effective powertrains. Optimizing energy use is crucial for extending range and reducing expenses. However, comparing inverter and motor efficiency based on inverter topologies is challenging due to biased methodologies that favor certain designs over others. This document introduces a novel optimization-based approach for enhancing partial load efficiency and minimizing chip area of single and dual traction inverters, indicating potential energy savings and cost reduction. Recent publications of both industry and academia underscore the importance of these design goals achieved by either novel inverter topologies or enhanced control methods. Two promising topologies with the inherent capability of partial load optimization are evaluated regarding chip area and system efficiency to find the most suitable concept for future electric vehicle power trains.

%\cite{martens2021loss}
\end{abstract}


%%%%%%%%%%%%%%%%%%%%%%%%%%%%%%%%%%% INTRODUCTION %%%%%%%%%%%%%%%
\section{Introduction}\label{sec:Intro}

The combination of modern wide bandgap (WBG) semiconductors, a DC-link voltage of 800\,V, and novel control algorithms encourages the use of specific power electronic topologies with certain motor configurations \cite{haring2023analysis,bhaskar2024}. This enables a significant boost in powertrain efficiency with a comparatively small increase in total inverter chip area. Improving system efficiency leads to a reduction in battery capacity size, which is the main cost driver for electric vehicles, whereas the chip area is the main cost driver for the inverter itself. Therefore, it is of great importance to consider the distribution of alternative chip area, efficiency, and circuit complexity ratios in power electronics converters with the aim of reducing overall vehicle costs and increasing range or decreasing battery size.

\begin{figure}[!t]
\vspace{-1em}
\begin{minipage}{0.48\textwidth}
\subfloat[\label{fig:B6}]{\includegraphics[height=0.4\textwidth]{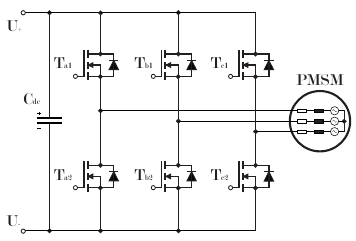}} \quad
\subfloat[\label{fig:T2}]{\includegraphics[height=0.4\textwidth]{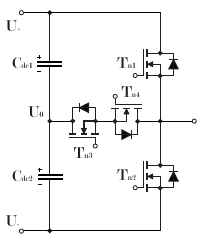}} \quad
\subfloat[\label{fig:B62y}]{\includegraphics[height=0.6\textwidth]{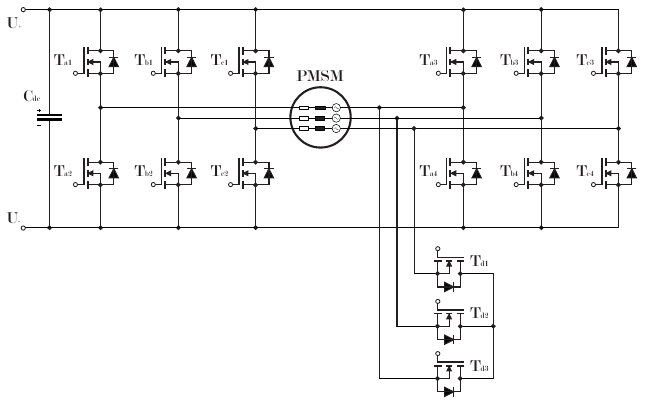}}
\caption{\label{fig:Topos}Schematic of inverter topologies: \textbf{(a)} B6 inverter in 3-leg form with closed star-point motor configuration, \textbf{(b)} TNPC inverter in 1-leg form and \textbf{(c)} B6$^2$-Y inverter in 3-leg form with open-winding motor configuration.}
\end{minipage}
\vspace{\distance}
\end{figure}

The multi-objective optimization process for the traction inverter comprises two distinct parts. The first ensures the inverter is capable of maintaining the full load over the entire operating range. The second part of the process is focused on reducing motor losses and inverter losses in the partial load area. The inverter losses consist of switching and conduction losses, which depend on the full load optimization design outcome of the semiconductor switches from the first phase. Fundamental losses are calculated using given datasets, while harmonic losses due to modulation are determined analytically with parameters fitted to measurements. The calculated inherent loss components are those of harmonic iron, magnet and copper losses. Finally, an overview of the impact of power loss and chip area on the aforementioned optimisation techniques is provided. This includes an investigation into the influence of the switching frequency, modulation strategy and space vector variation of the topologies under consideration. The Worldwide Harmonized Light-Duty Vehicles Test Procedure (WLTP) serves as the reference mission profile for evaluating efficiency in the partial load area. In essence, the optimization tool facilitates a comprehensive evaluation of the impact of various inverter concepts on desired powertrain attributes. 

Fig. \ref{fig:Topos} illustrates three distinct inverter topologies: \textbf{(a)} B6, \textbf{(b)} TNPC, and \textbf{(c)} B6$^2$-Y. The widely-used B6 inverter serves as a benchmark, renowned for its cost-effectiveness, well-known operation, and control simplicity. The TNPC inverter is shown as a single leg in subfigure \textbf{(b)} and is evaluated as being connected with a star-point motor configuration. The TNPC inverter operates in a three-level (3L) mode by controlling all switches, thereby offering several advantages. One such advantage is the reduction of harmonic content on the motor windings. Alternatively, it can function as a two-level (2L) inverter by utilizing only the upper and lower switches, $T_{n1}$ and $T_{n2}$, a feature that is often overlooked. This flexibility allows the design of multiple TNPC inverter types with varying chip areas and limited 3L capability over the torque-speed map of the motor. As an example, at high torque and speed operation points, the system could use solely the main switches $T_{n1}$ and $T_{n2}$ for 2L operation and employ all switches for 3L operation mainly during the partial load operation. This enables an efficiency increase during the reference efficiency mission profile with a decreased chip area of the switches $T_{n3}$ and $T_{n4}$, as low torque operation points do not result in tremendous currents and excessive junction temperatures in the mid-point switches. Moreover, the choice of zero space vectors is crucial because the use of alternative vectors affects the power loss ratios between the inner switches ($T_{n3,4}$) and outer switches ($T_{n1,2}$) of the TNPC topology. This choice can reduce the chip size of the inner switches, but it increases the total switching losses. TNPC$_A$ uses all zero space vectors and represents the standard space vector configuration. On the other hand, TNPC$_B$ uses only the [1,1,1] space vector state, while TNPC$_C$ uses both [1,1,1] and [-1,-1,-1] states and chooses the best one depending on minimal switching losses. These different modulation strategies directly influence chip area and efficiency outcomes, optimizing the 2L-/3L- capability of types B and C to achieve smaller total chip areas by avoiding overloading of the mid-point switches.

\begin{figure}[]
\begin{minipage}{0.48\textwidth}
    \centering
    \includegraphics[width=\textwidth]{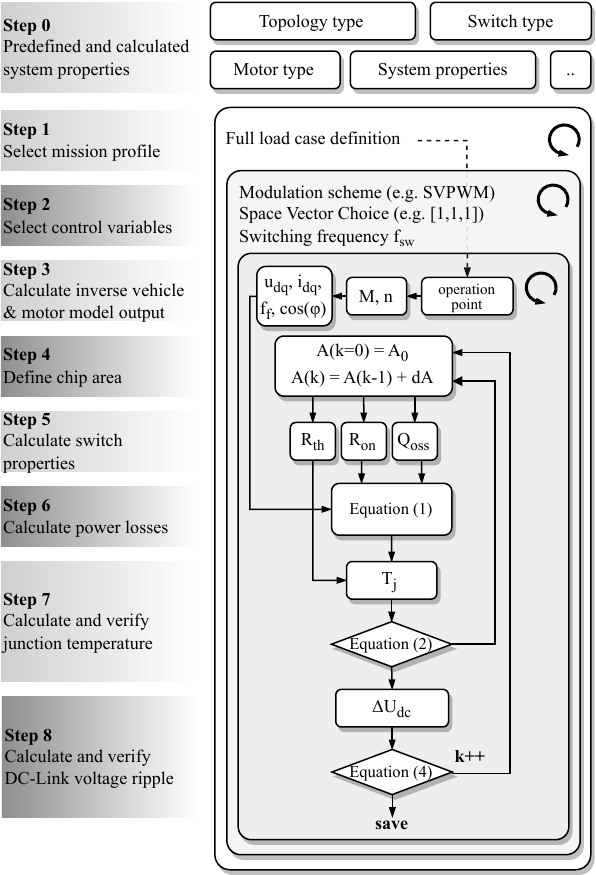}
    \caption{Processing and optimization framework for full load operation to determine minimum chip area requirements and optimal switching frequency while meeting design constraints, including DC-link ripple voltage and maximum junction temperature. \label{fig:Metrik}}
\end{minipage}
\vspace{\distance}
\end{figure}

The B6$^2$-Y inverter, a novel design introduced by Hyundai Motor Company in 2023, combines star-point and H-bridge motor operation. It is designed for motors with open-winding configurations, allowing it to utilize the full DC-link voltage instead of $V_{dc}/\sqrt{3}$, assuming common mode injection for star-point motor configurations. Furthermore, it enables the independent design of both B6 bridges, including the integration of WBG components such as SiC MOSFETs as well as conventional semiconductors like Si IGBTs. This shifts the focus of design from efficiency optimization to cost compensation. The open-winding motor operation with both B6 bridges reduces the needed maximum phase currents compared to a conventional motor in a star-point configuration utilizing the same DC-Link voltage for all load points. This operation is called dual-mode (H-mode) and is achieved by controlling all switches except for $T_{d1}$, $T_{d2}$, and $T_{d3}$. However, this leads to higher switching losses and, depending on the modulation strategy, can cause dominant harmonic voltage ripples and increase harmonic motor losses. To address this, solely the left B6 bridge of the inverter is used during lower output power operation points. This so-called star-point mode (Y-mode) includes the switches $T_{a1}$, $T_{a2}$, $T_{b1}$, $T_{b2}$, $T_{c1}$, and $T_{c2}$, creating a switchable star-point with three additional switches $T_{d1}$, $T_{d2}$, and $T_{d3}$. The B6$^2$-Y$_A$ inverter operates in H-mode with inversely actuated switches, meaning states $T_{n1} = T_{n4}$ and $T_{n2} = T_{n3}$. In contrast, B6$^2$-Y$_B$ operates in H-mode using inverse input signals for the left and right inverter halves $m_{12} = \hat{M} \cdot sin (wt + \varphi)$ and $m_{34}= -m_{12}$ as done in \cite{antivachis2020analysis} with the unipolar modulation. In this work, the B6 and TNPC topologies operate using 2L- and 3L-SVPWM, while the B6$^2$-Y topology uses SPWM in H-mode and SVPWM in Y-mode.

%%%%%%%%%%%%%%%%%%%%%%%%%%%%%%%%%%% SECTION 01 %%%%%%%%%%%%%%%%%
\section{full load Operation}
\label{sec:flo}
The load limit of the switches is defined by the full load operation, as this is where the highest switch power losses occur. The switches used for evaluation are 4$^{th}$ generation SiC MOSFETs (1200\,V \& 750\,V) from ROHM Co., Ltd. \cite{ROHM_datasheet}. A 300\,kW Permanent Magnet Synchronous Motor (PMSM) in \textit{48/8} configuration is used for operation and loss analysis as the standard motor for the B6 and TNPC topologies. A scaled motor model that yields equal mechanical operation output has been designed for the B6$^2$-Y topology. The motor model is based on equal machine data, but the turn number is increased by $7/4 \approx \sqrt{3}$, leading to decreased phase currents in (H)-operation to generate the same output power.

Fig. \ref{fig:Metrik} illustrates the simplified calculation steps for the results of full load operation, which are detailed in the following subsections. The process begins with step 1 by determining a specific full load point, which is defined by the boundary of the speed torque map of the motor. In step 2, input parameters are set, including the modulation scheme, space vector choice, and switching frequency $f_{sw}$. A vehicle model with the parameters specified in Table \ref{table:vehicle_paramssection} is employed to derive the motor torque and speed values. Designated voltages $u_q$, $u_d$, currents $i_q$, $i_d$, motor fundamental frequency $f_f$, and the power factor $cos(\varphi)$ are determined based on look-up tables from test bench measurements. Right after alternating the individual switch chip area in step 3, the inherent system properties, such as thermal resistance $R_{th}$, on-state resistance $R_{on}$, and output capacitance charge $Q_{oss}$, are calculated within step 5. It is crucial to ensure that thermal and electrical system properties remain within acceptable limits. This verification occurs in subsequent steps for the DC-link voltage ripple and the individual junction temperature of all switches. If these properties are not within specific limits, the simulation steps 5 to 8 are repeated with an adapted chip area $A(k) = A(k-1) + dA$. 

\begin{figure}[]
    \centering
    \includegraphics[width=0.5\textwidth]{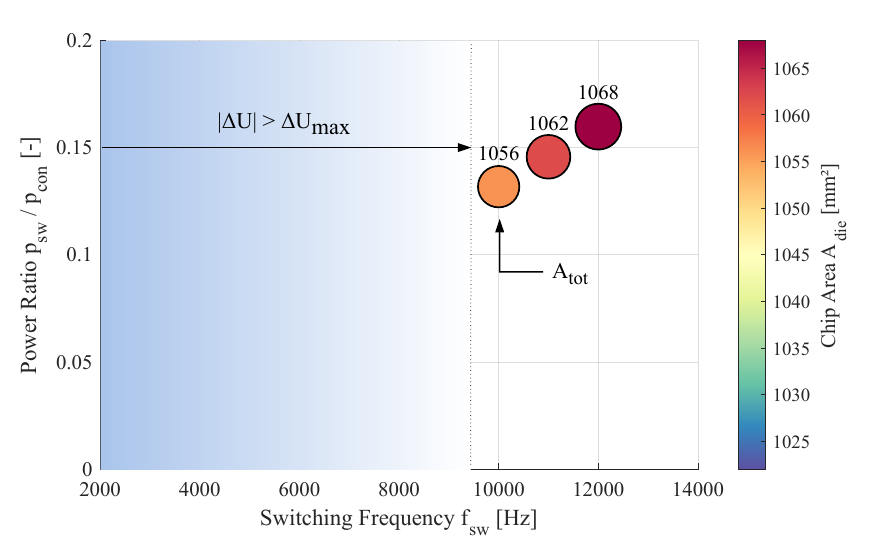}
    \caption{B6 topology chip area $A_{die} = \sum A_{die,j}(n_i,M_i,f_{sw})$ as a function of switching frequency $f_{sw}$ under the defined full load constraints $T_{j,max}$ and $\Delta U_{max}$ at $M = 589$\,Nm and $n = 5000$\,rpm.}
    \label{fig:FL_Adiefsw}
\vspace{\distance}
\end{figure}

\begin{table*}[!b]
    \caption{Longtudinal vehicle model parameters for calculating the motor torque-speed map.}
    \label{table:vehicle_paramssection}
    \centering
    \renewcommand{\arraystretch}{1.2} % Adjust vertical spacing
    \begin{tabular}{l*{12}{c}}
        \hline 
        $A$ & $C_d$ & $\rho_{\text{air}}$ & $R_r$ & $g$ & $R_w$ & $m$ & $G$ & $\eta_{\text{gear}}$ & $J_{\text{axle}}$ & $J_{\text{EM}}$ & $N_{\text{EM}}$ \\
        {[}m\textsuperscript{2}{]} & {[-]} & {[}kg/m\textsuperscript{3}{]} & {[-]} & {[}m/s\textsuperscript{2}{]} & {[}m{]} & {[}kg{]} & {[-]} & {[-]} & {[}kg$\cdot$m\textsuperscript{2}{]} & {[}kg$\cdot$m\textsuperscript{2}{]} & {[-]} \\
        \hline
        2.22 & 0.25 & 1.25 & 0.01 & 9.81 & 0.345 & 1927 & 12.4 & 1 & 0 & 0 & 1 \\
        \hline 
    \end{tabular}
    \vspace{-10pt}
\end{table*}

\subsection{Inverter Power Losses}

The inverter loss $P_{inv}$ is calculated by summing up the losses during one fundamental period $T_0 = \frac{1}{f_f}$ for each switch $j$ in the topology, defined by Equation (\ref{eq:inv}). The conduction loss component $p_{con}$ is linearly related to the drain-to-source resistance $R_{DS,on}$, resulting in conduction loss scaling inversely with the chip area $P_{con} \propto R_{DS,on} \propto 1/A_{die}$. The switching losses $P_{sw}$ are determined by accounting for energy dissipation of the switches as described in \cite{friedli2009semiconductor}. The output capacitance charge $Q_{oss}$-related turn-on losses \cite{deboy2017perspective} of the MOSFETs are considered, as their contribution to switching losses becomes dominant at certain partial load points.

\begin{equation}
\label{eq:inv}
P_{inv} = \sum_{j=1}^{N} P_{mos_j} = \sum_{j=1}^{N} \left[ P_{sw_j} + P_{con_j} \right]
\end{equation}

\begin{figure}[]
    \centering
    \includegraphics[width=0.5\textwidth]{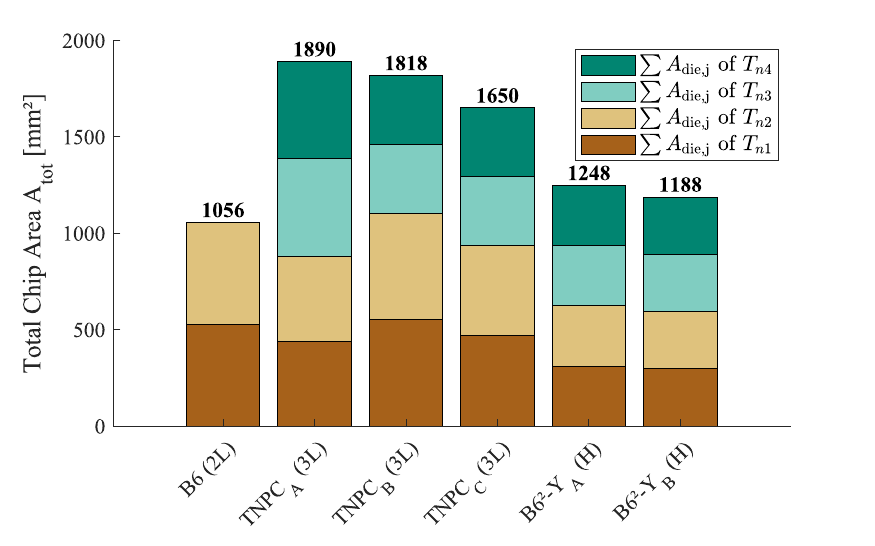}
    \caption{The total chip area $A_{tot} = max\left \{ A_{die}(n_i,M_i,f_{sw}) \right \}$ required to meet the full load operation constraints $T_{j,max}$ and $\Delta U_{max}$ in all operation points.}
    \label{fig:FL_bar_chart}
\vspace{\distance}
\end{figure}

\subsection{Operation Constraints}

The junction temperature of each switch is calculated using a heat balance model \cite{friedli2009semiconductor}, as shown in Equation (\ref{eq_TjAdie}). Each switch is thermally isolated from each other and has a maximum junction temperature of $T_{j,\text{max}} = 175 ^\circ \text{C}$. The thermal resistance $R_{th}$ between the chip and the heat sink ($T_{hs} = 65 ^\circ \text{C}$) temperature regulated by the cooling fluid is described by a mean fit based on empirically measured values of an average cooling system for traction inverters, as listed in Equation (\ref{eq:Rth}). 

\begin{equation}
{T_{\text{j}}} = {T_{{\text{hs}}}} + {R_{{\text{th}}}} \cdot {P_{{\text{mos}}}} \leq T_{j,\text{max}}
\label{eq_TjAdie}
\end{equation}

\begin{equation}
{R_{{\text{th,j}}}}_{\left[ \frac{\text{K}}{\text{W}} \right]} = 3 \cdot A_{\text{die,j }  \left[ \text{mm}^2 \right]} ^{ - 0.4}
\label{eq:Rth}
\end{equation}

The total chip area is divided into individual chips, each with a granularity of approximately 25\,mm². Excess temperature due to short-term overcurrents caused by an active short-circuit state are avoided by optimized control of the switches, and therefore does not lead to additional chip area for the topologies considered herein.

It is essential to consider the charge balance $Q_{ac}$ of the DC-link capacitor. This is crucial for guaranteeing the proper functioning of other components connected to the battery, as excessive voltage ripples could impede their operation. To ensure this, the voltage ripple $\Delta U$ across the capacitor must be kept below the limit as defined in Equation (\ref{eq:deltaU}). A capacitance of $C_{dc}=500 \mu$F is employed, with the TNPC capacitance exhibiting an identical total energy for the split capacitors. Furthermore, it is important for the TNPC topology to consider the voltage balancing of the two capacitors, which can be effectively managed by advanced balancing algorithms \cite{stamer2018investigation}.

\begin{equation}
\Delta U = \left| \frac{Q_{\text{ac}}}{C_{\text{dc}}} \right| = 
\left| \frac{\int i_{\text{c}} \cdot dt}{C_{\text{dc}}} \right|
 \leq \Delta U_{\text{max}} = 15  \text{V}
\label{eq:deltaU}
\end{equation}
In Fig. \ref{fig:FL_Adiefsw}, the minimum total chip area required for the B6 inverter at full load is illustrated. Each bubble represents the minimum chip area at a specific switching frequency needed to meet the junction temperature constraint. The $\Delta$U arrow indicates whether the ripple voltage constraint is satisfied, which is true for $f_{sw} >$ 9.5\,kHz at the worst-case maximum load. Additionally, the switching to conduction loss ratio $p_{sw}/p_{con}$ on the y-axis shows that with increasing switching frequency, both the required chip area and the ratio increase. The required chip area is significantly influenced by the modulation scheme, space vector choice, and the specific full load case.

To summarize this, Fig. \ref{fig:FL_bar_chart} presents a comparison of the minimum total chip areas $A_{tot}$ of the topologies based on all full load operation cases and the fulfillment of both constraints. It is clear that the TNPC modulation schemes A, B, and C differ in terms of required chip area. The difference can be attributed to the fact that the standard modulation scheme results in significant losses in the mid-point switches, namely $T_{3}$ and $T_{4}$, at the maximum torque and minimum speed operation points.

%%%%%%%%%%%%%%%%%%%%%%%%%%%%%%%%%%% SECTION 02 %%%%%%%%%%%%%%%%%
\section{Partial Load Operation}
\label{chap:curr_driven_rect}
\begin{figure}[]
\begin{minipage}{0.48\textwidth}
    \centering
    \includegraphics[width=\textwidth]{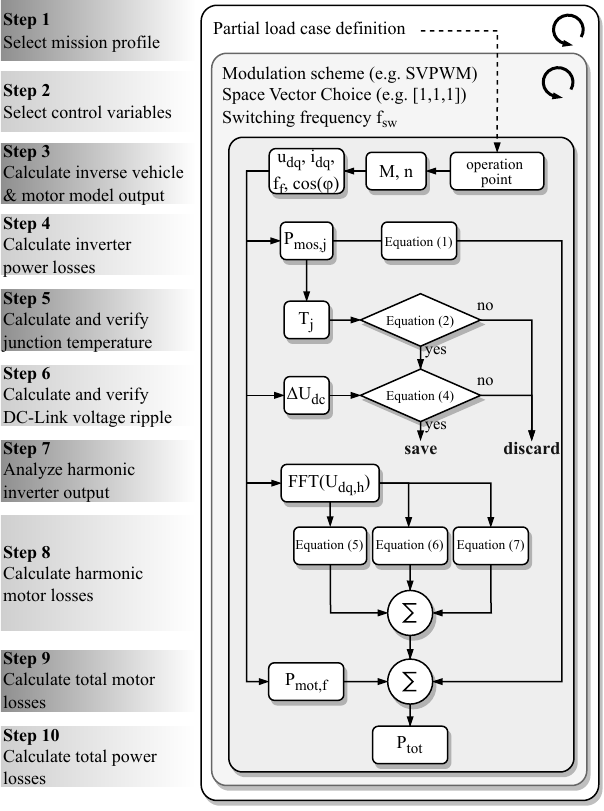}
  \caption{  \label{fig:Metrik_PL}Processing and optimization framework for partial load operation to determine total power losses depending on various modulation variables and optimal switching frequency while meeting design constraints, including allowable DC-link ripple voltage and maximum junction temperature.}
\end{minipage}
\vspace{\distance}
\end{figure}

The efficiency of the system under partial load operation is assessed using WLTP-defined operation points. The evaluation involves calculating total system losses, including harmonic motor losses caused by high-frequency ripple voltage across the motor windings. The pre-defined chip areas of all switches by the full load optimization framework are used for the partial load optimization steps, illustrated in Fig. \ref{fig:Metrik_PL}. Steps 1 to 6 are identical to the steps in Fig. \ref{fig:Metrik}. Changing the control variables leads to the decision of whether the partial load point are fed by the inverter or alternative operation modes are required, e.g., if an increased switching frequency is necessary. Step 7 involves analyzing the harmonic content of the inverter output $U_{dq,h}$. This step helps identify and mitigate harmonic distortions that affect the harmonic motor losses, which are calculated in step 8. Within step 9, the total motor losses are calculated, combining both the fundamental and harmonic loss parts. Finally, step 10 calculates the total power losses of the system. And therefore, it provides a holistic view of the system's energy efficiency with a predefined chip area.

\subsection{Harmonic Motor Losses}

The harmonic voltage ripple across the motor windings is divided into d- and q-axis components and transformed to the frequency spectrum using Fast Fourier Transformation (FFT), which allows for the identification of individual harmonic components $U_{d,h}(f_{h})$, and $U_{q,h}(f_{h})$. These harmonics are used with specific harmonic motor parameters, including the d- and q-axis motor inductances $L_{d,h}(f_{h})$, and $L_{q,h}(f_{h})$. Harmonic iron, magnet and copper loss factors $R_{iron,h}(f_{h})$, $R_{mag,h}(f_{h})$, and $R_{cu,h}(f_{h})$ are used, whereby the first two do not directly represent ohmic resistances. Instead, they indicate an equivalent resistive loss due to eddy currents. The scaling factors $k_{iron}$, $k_{mag}$, and $k_{cu}$ are used based on an analytical model fitted to motor measurement data. It is important to remember that these variables, parameters, and look-up tables differ depending on the utilized motor configuration and the target load points. For the frequency components under consideration, the minimum and maximum frequencies $f_{min}=f_{sw}/2$ and $f_{max} =$ 1\,MHz were selected, as this results in taking into account nearly all harmonic loss parts. The harmonic losses are calculated in accordance with the Equations (\ref{eq:pharmcu}),\,(\ref{eq:pharmiron})\,and\,(\ref{eq:pharmmagnet}).

\begin{equation}
\label{eq:pharmcu}
P_{cu,h} = k_{{cu}} \cdot \sum_{f_{{min}}}^{f_{{max}}} \left[ \frac{R_{\text{cu},h}}{f_h^2} \cdot \left( \frac{U_{d,h}^2}{L_{d,h}^2} + \frac{U_{q,h}^2}{L_{q,h}^2} \right) \right]
\end{equation}

\begin{equation}
\label{eq:pharmiron}
P_{{iron},h} = k_{{iron}} \cdot \sum_{f_{{min}}}^{f_{{max}}} \left[ \frac{U_{d,h}^2+U_{q,h}^2}{R_{\text{iron},h}} \right]
\end{equation}

\begin{equation}
\label{eq:pharmmagnet}
P_{{mag},h} = k_{{mag}} \cdot \sum_{f_{{min}}}^{f_{{max}}} \left[ \frac{U_{d,h}^2}{R_{\text{mag},h}} \right]
\end{equation}

The harmonic motor losses are calculated by summing up the individual harmonic loss components.

\begin{equation}
\label{eq:Pmoth}
P_{{mot,h}} = P_{{mag,h}} + P_{{iron,h}} + P_{{cu,h}}
\end{equation}

\begin{figure}[]
    \centering
    \includegraphics[width=0.5\textwidth]{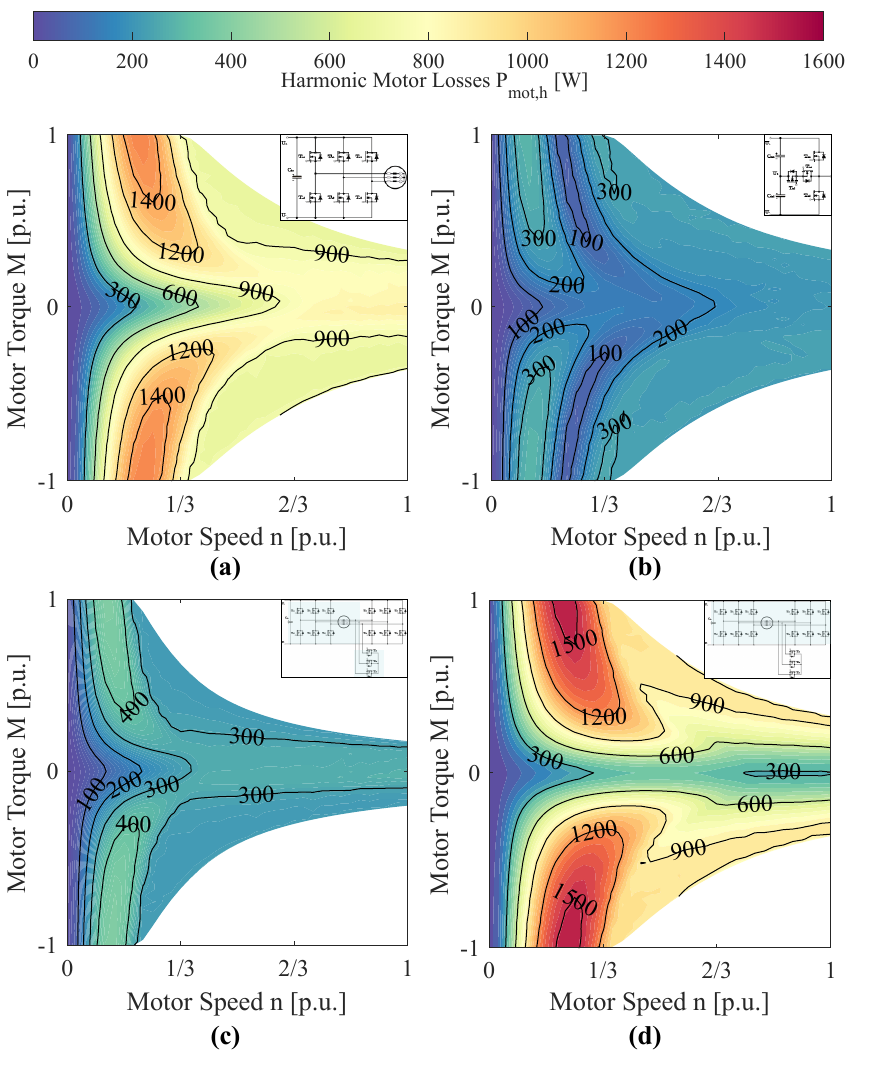}
    \caption{Harmonic motor losses $P_{mot,h}$ over the motor torque speed map at 10 kHz illustrated for \textbf{(a)} B6 topology with 2L-SVPWM and PMSM in star-connection, \textbf{(b)} TNPC$_A$ topology with 3L-SVPWM, \textbf{(c)} B6$^2$-Y topology in Y-mode with SVPWM, and \textbf{(d)} B6$^2$-Y$_B$ topology in H-mode with SPWM.}
    \label{fig:PL_Pmoth}
\vspace{\distance}
\end{figure}

Fig. \ref{fig:PL_Pmoth} depicts the harmonic motor losses $P_{mot,h}$ of the evaluated topologies with their corresponding motor configurations at $f_{sw}=$10\,kHz. The graphs are normalized regarding the motor torque and speed axes. Subfigure \textbf{(a)} illustrates the harmonic loss components of the motor when a B6 topology is used. The principal losses due to the 2L operation emerge at high torque values within the base speed range. The TNPC$_{A}$ inverter \textbf{(b)}, on the other hand, reduces harmonic losses due to its 3L operation in the base speed area, creating harmonic motor minima where the 2L operation has maxima. This is especially beneficial for the efficiency increase in the partial load area. Theoretical details regarding the harmonic voltage distortion shift due to multi-level operation are comprehensively covered in \cite{ruderman2010pwm}. The B6$^2$-Y$_A$ topology in Y-mode \textbf{(c)} is characterized by a different motor speed torque map shape compared to its counterpart in H-mode \textbf{(d)}. The output power limit is caused by the decreased voltage $V_{dc}/\sqrt{3}$, considering common mode injection. The harmonic content of the Y-mode differs from that of the H-mode due to the lower voltage amplitude, which significantly reduces the total harmonic power loss of the system. This highlights the system's efficiency advantages due to partial load optimization. Subfigure \textbf{(d)} shows that the B6$^2$-Y$_B$ topology in H-mode can supply all full load cases defined by the speed torque map outline, resulting in similar losses at high torque and slightly lower losses at low torque and high speed compared to the B6 topology in SVPWM due to utilizing the full DC-Link voltage using modulation strategy B. The shape compared to subfigure \textbf{(a)} differs mostly due to not using the freewheeling condition by activating e.g. $T_{a1}$ and $T_{a3}$.

%This meets expectations, as the fundamental coil equation $di_H/dt_H=U_H/L_H \approx \sqrt{3 U_Y}/(3 L_Y)$ and the maximum current ripple can be approximated with $R_H \cdot i^2_H \approx 3 \cdot R_Y \cdot 1/3 \cdot i^2_Y$. This results in slightly smaller copper losses for the B6$^2$-Y topology due to a slimmer coil diameter and therefore smaller skin and proximity loss components. 

%%%%%%%%%%%%%%%%%%%%%%%%%% hier weitermachen mit spell check

\subsection{System Power Loss Evaluation}

The total system loss $P_{{tot}}$ depends on inverter losses $P_{{inv}}$, fundamental motor losses $P_{{mot,f}}$, and harmonic motor losses $P_{{mot,h}}$, which are summed up in Equation (\ref{eq:Ptot}). The harmonic motor outputs of the standard PMSM with star-point and the open-winding motor in H- and Y-operation differ fundamentally from each other depending on the modulation strategy. The fundamental losses increase in Y-mode in the field weakening range as less voltage is available. Lower modulation losses are therefore offset by higher fundamental losses.

\begin{equation}
\label{eq:Ptot}
P_{{tot}} = P_{{inv}} + P_{{mot,h}} + P_{{mot,f}}
\end{equation}

The energy loss per 100 kilometers $\Delta e$ serves as a vivid comparison metric, determined by evaluating the total power loss $P_{{tot}}$ over one WLTP cycle and scaling it by the mission profile track distance $s_{{cycle}}$ to 100\,km.

\begin{equation}
\begin{split}
\Delta e &= E_{{tot,cycle}} \cdot \left[ \frac{100\text{\,km}}{s_{{cycle}}} \right] \\
&= \int_{0}^{T_{{cycle}}} \left( P_{{tot}} \cdot dt \right) \cdot \left[ \frac{100 \text{\,km}}{\int_{0}^{T_{{cycle}}} v_{{cycle}} \cdot dt} \right]
\end{split}
\end{equation}

The total mean partial load losses are calculated by integrating the total system losses $P_{tot}$ over the complete mission profile and dividing them by with the required total time of the drive cycle.

\begin{equation}
    {P_{tot,m}} = \frac{1}{T_{{cycle}}}\int_{0}^{T_{{cycle}}} P_{{tot}} \cdot dt 
\end{equation}

% %%%%%%%%%%%%%%%%%%%%%%%%%%%%%%%%%%% SECTION 03 %%%%%%%%%%%%%%%

\subsection{Advanced Control Strategies}

\begin{figure}[]
    \centering
    \includegraphics[width=0.5\textwidth]{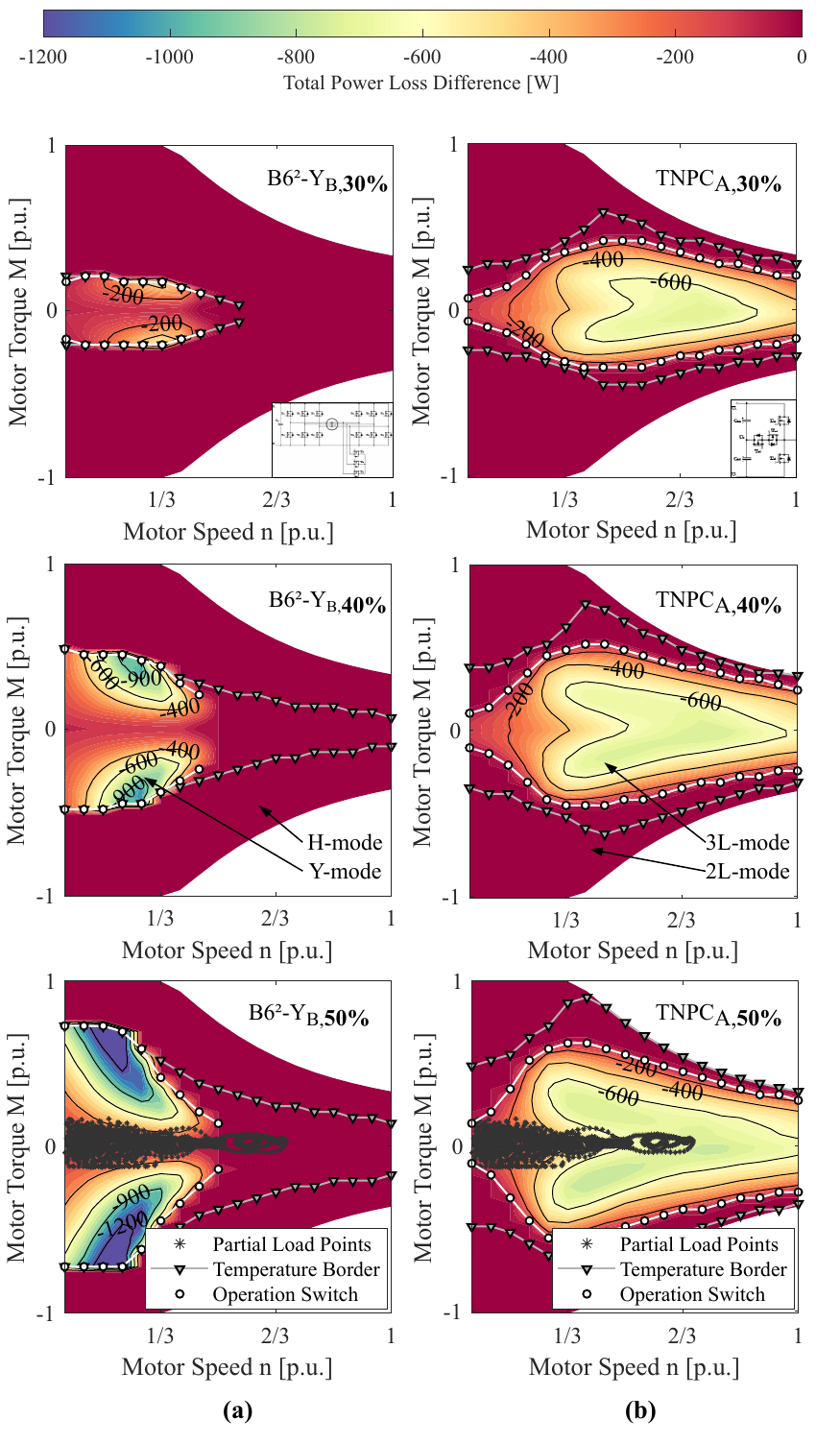}
    \caption{Operation boundaries, temperature limits, and total loss differences of chip area variable B6$^2$-Y$_B$ inverter designs \textbf{(a)} on the left and TNPC$_A$ inverter designs \textbf{(b)} on the right at $f_{\text{sw}} = 10 \, \text{kHz}$. The gray triangle line represents the thermal boundary of the 3-level and Y-operation, whereas the white dotted line represents the power loss equilibrium of both modes. The black stars in the lower plots show the partial load points based on the WLTP as a reference.}
    \label{fig:PL_opswitch}
\vspace{\distance}
\end{figure}

The total system loss $P_{{tot}}$ can be reduced by operating the power electronics converters in alternative modes, such as utilizing 3-level operation for the TNPC inverter. However, the use of the 3L-mode is restricted by the chip area of switches $T_{n3}$ and $T_{n4}$, as smaller semiconductor areas cannot handle full load currents. Similarly, an efficiency boost is enabled by the Y-mode of the B6$^2$-Y topology, which is limited by the chip area of the switches $T_{d1}$, $T_{d2}$, and $T_{d3}$. Fig. \ref{fig:PL_opswitch} illustrates various chip area variable converter designs and their operational boundaries. It emphasizes that chip area not only defines operational limits, but also significantly influences losses in both the 3L-mode and Y-mode of the TNPC and the B6$^2$-Y topology. Chip area usage is defined based on the B6 total chip area as a reference. For example, a TNPC$_{A,30\%}$ inverter design has an increased chip area of 30\,\% compared to the B6 inverter, leading to a total chip area of $1.3 \cdot 1056 \, \text{mm}^2 \approx 1373 \, \text{mm}^2$.

The torque and speed-dependent operation boundaries of the chip area variable converters are highlighted by the gray triangle line, showing the maximum operational capability of each design in the 3L- and Y-mode. In contrast, the operation boundaries defined by the circled line are determined by the power loss balance of different modes and are used to evaluate the minimum power losses of the systems. Both the operation boundaries and temperature limits vary significantly between the topologies. The B6$^2$-Y$_B$ topology \textbf{(a)} requires 13\,\% more chip area to enable the H-mode itself, whereas the TNPC$_A$ topology \textbf{(b)} starts with an equivalent chip area as the B6 inverter, and any additional area directly converts into increasing the operation boundary. Furthermore, the locations of the minimum losses and the movement of the operation boundaries with increasing chip areas are fundamentally different. Generally, the operation boundary of the B6$^2$-Y$_B$ inverter lies outside the field weakening area of the motor because in Y-mode, the field weakening current $i_d$ increases significantly to achieve the desired maximum speed levels, resulting in dominant fundamental copper losses. On the other hand, the temperature boundary of the $TNPC_A$ inverter initially increases within the field weakening area but maintains a distance from the corners of absolute maximum torque. This is due to the overutilization of the mid-point switches $T_{n3}$ and $T_{n4}$ due to the frequent use of [0,0,0] zero space vectors in this region. This highlights the advantage of alternative TNPC modulators B and C, which enable zero space vector utilization using the main switches $T_{n1}$ and $T_{n2}$ that already possess a high chip area to support full load operation.

The maximum loss difference of the B6$^2$-Y$_B$ topology is significantly higher compared to the TNPC$_A$ inverter, but does not result in lower total losses considering the partial load point distribution across the torque-speed map. The B6$^2$-Y$_B$ inverter fails to cover the entire WLTP with the Y-mode even with a 50\,\% increase in chip area size and does not exhibit superior loss differences for |M| < 100\,Nm. Additional chip area does not alter this general phenomenon. It is noteworthy that operating the B6$^2$-Y inverter with modulator A results in increased harmonic copper and magnetic losses due to the absence of the freewheeling condition, which in turn leads to 0-voltages across the motor windings.  

%The color shift throughout the partial load area due to chip area increase directly converts into an partial load energy loss decrease $\Delta e$.

% %%%%%%%%%%%%%%%%%%%%%%%%%%%%%%%%%%% SECTION 04 %%%%%%%%%%%%%%%

\section{System Evaluation and Comparison}

\begin{figure}[h]
    \centering
    \includegraphics[width=0.5\textwidth]{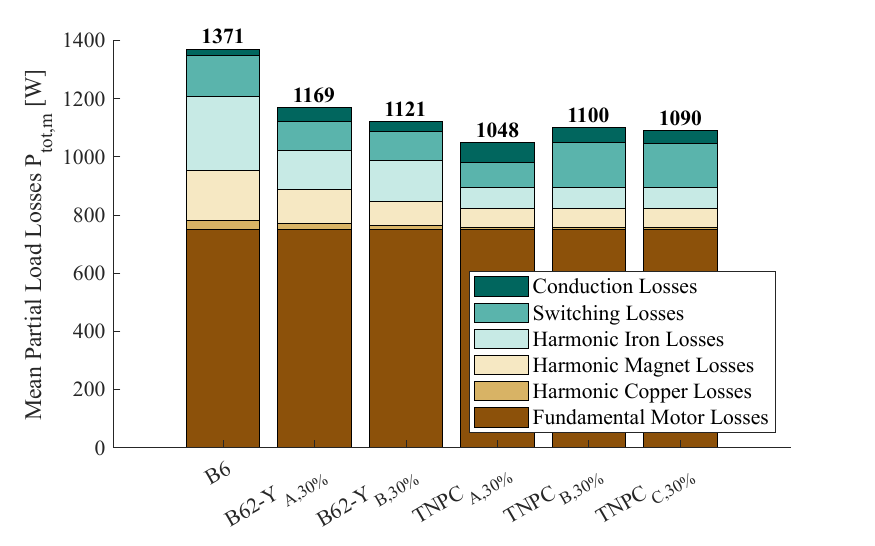}
    \caption{Individual partial load power loss components of various inverter topology designs have a 30\,\% additional chip area size compared to the B6 inverter. The loss components are subdivided into fundamental motor losses, harmonic motor loss components, switching losses and conduction losses at a switching frequency $f_{sw}=$ 10\,kHz.}
    \label{fig:barchart_ploss}
\end{figure}

Fig. \ref{fig:barchart_ploss} shows the individual mean partial load loss components for several inverter topology designs, incorporating a 30\,\% additional chip area compared to the B6 inverter. The inverter topology designs examined include B6, B6$^2$-Y with modulation types A and B, and TNPC with the modulation types A, B, and C. The loss components are subdivide into conduction and switching losses for the inverters, as well as harmonic and fundamental losses of the motor. The results show that the B6 inverter exhibits the highest mean partial load power losses of 1371\,W, mainly caused by very high harmonic motor losses.  The B6 inverter is followed by the B6$^2$-Y$_A$ design with 1169\,W due to the partial use of the Y-mode. The design has generally lower harmonic motor losses due to the switchable motor configuration, but increased conduction losses due to the additional three switches $T_{d1}$, $T_{d2}$, and $T_{d3}$. The switching losses are decreased due to lower chip areas of the left main switches $T_{a1}$, $T_{a2}$, $T_{b1}$, $T_{b2}$, $T_{c1}$, and $T_{c2}$, whereas the additional three switches only switch if the configuration is changed between Y and H. The harmonic motor losses of the B6$^2$-Y$_B$ are 48\,W smaller due to improved harmonic motor losses. This improvement is achieved using the freewheeling state of the topology in H-mode for operation points not covered by the Y-mode, as indicated by the operation switch for optimal efficiency in Fig. \ref{fig:PL_opswitch}. The TNPC$_A$ design leads to the lowest mean partial load losses, due to covering most of the partial load points and reduced harmonic motor losses. The observed differences in power loss between the 3L TNPC inverters can be attributed to the fact that only half of the DC-link voltage is switched and the resulting differences in switching losses. The advantage of modulator B and C of increased 3L-area at same chip area does not enhance the efficiency because at 30\,\% additional chip area size, even TNPC$_A$ covers most of the partial load points.

The optimal switching frequency calculation is based on an evaluation of the system parameters $T_j$, $\Delta U_{dc}$ and a minimum search function considering the partial load point $(n_i,M_i)$, the control strategy, modulator $mod$ and the space vector choice $svc$:

\begin{equation}
    f_{sw,opt} = min\left \{     P_{tot}(n_i,M_i,mod,svc,..)
 \right \}
\end{equation}

The switching frequency range of different inverter topology designs varies, as the optimal switching frequency $f_{sw,opt}$ depends on the harmonic motor losses and the total switching losses of the inverter at each operation point. An average switching frequency of 10 to 13\,kHz is observed, whereas the minimum values of $f_{sw,opt}$ vary between 6 to 7\,kHz, mainly depending on the voltage ripple constraint. The maximum values of $f_{sw,opt}$ for the TNPC and B6$^2$-Y designs vary between 16 and 18\,kHz. Impact factors for the switching losses are the chip sizes of the controlled and uncontrolled switches. This analysis doesn't consider parasitic capacitance motor losses, which increase with switching frequency. Therefore, including those losses could limit the maximum sensible switching frequency further.

\begin{table}[]
\vspace{0.5em}
\centering
\caption{Overview of inverter types illustrated in Fig. \ref{fig:pareto} and their utilized operation modes inclusive modulation types and 3L-zero space vectors for the TNPC variants.}
\label{table:paretotypes}
\begin{tabular}{lccc}
\hline 
{{Inverter type}} & {{Operation modes}} & {{3L-zero vectors}} \\
\hline 

B6 & 2L-mode & -  \\
B6²-Y$_{A}$(H) & H-mode (A) & -  \\
B6²-Y$_{B}$(H) & H-mode (B) & -  \\
TNPC$_{A}$(3L) & 3L-mode & unrestricted  \\
TNPC$_{B}$(3L) & 3L-mode & [1,1,1] \\
TNPC$_{C}$(3L) & 3L-mode & [1,1,1] \& [-1,-1,-1] \\
 & & \\
B6²-Y$_{A}$(H/Y) & H- (A) \& Y-mode & -  \\
B6²-Y$_{B}$(H/Y) & H- (B) \& Y-mode & -  \\
TNPC$_{A}$(2L/3L) & 2L- \& 3L-mode & unrestricted  \\
TNPC$_{B}$(2L/3L) & 2L- \& 3L-mode & [1,1,1] \\
TNPC$_{C}$(2L/3L) & 2L- \& 3L-mode & [1,1,1] \& [-1,-1,-1] \\
\hline 
\end{tabular}
\vspace{\distance}
\end{table}

\begin{figure}[]
\vspace{-1em}
    \centering
    \includegraphics[width=0.5\textwidth]{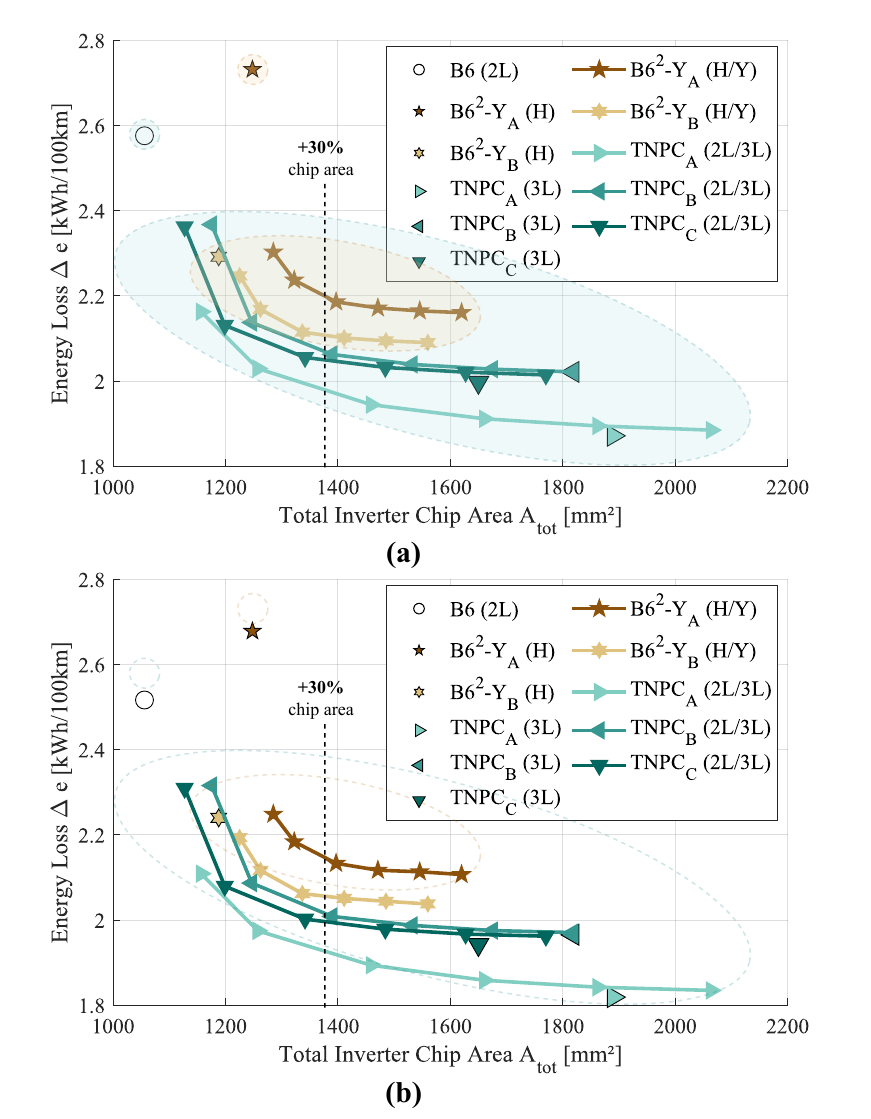}
    \caption{Energy loss $\Delta e$ and total inverter chip area $A_{die}$ Pareto fronts of the evaluated inverter topologies with either 2L-, 3L-, 2L/3L-mode for the TNPC topology and Y-mode, H-mode or Y/H-mode for the B6$^2$-Y topology. The same switching frequencies of \textbf{(a)} 10\,kHz and \textbf{(b)} $f_{sw,opt}$ are used.}
    \label{fig:pareto}
\vspace{\distance}
\end{figure}

Fig. \ref{fig:pareto} illustrates the Pareto fronts of the topologies in different optimization variants considering chip area at a switching frequency \textbf{(a)} $f_{\text{sw}} = 10 \, \text{kHz}$ and \textbf{(b)} $f_{\text{sw,opt}}$. Table \ref{table:paretotypes} summarizes the employed operation modes and 3L-zero space vector restrictions for the evaluated inverter designs. The use of 3L- for 2L-mode and the implementation of Y- in place of H-mode result in reduced energy losses $\Delta e$ for both inverter types, although this entailes disparate increases in chip area. In terms of chip area and system losses, the TNPC inverter exhibited superior performance relative to the B6$^2$-Y inverter for the modulation strategies and motor types employed in this study. In addition to the influence of modulation strategies on chip area and total losses, it should be noted that these also influence the chip area required to achieve full load 3L capability. This is indicated in the lines by the right-hand end point, defining the maximum chip area. TNPC$_B$ and TNPC$_C$ are operated in the partial load area the same way as the TNPC$_A$, as explained in section \ref{chap:curr_driven_rect}.  The single dots of the 3L TNPC inverters for modulators A and C do not cross the lines of the 2L/3L designs because the 2L/3L inverters start with additional chip area for the upper and lower switches $T_{n1}$, and $T_{n2}$ which is not the case for the modulator B, as illustrated in Fig. \ref{fig:FL_bar_chart}. For instance, the TNPC$_A$(3L) design is not capable of the 2L-mode over the complete operation area due to the low chip area of the main switches $T_{n1}$ and $T_{n2}$ as mentioned in Table \ref{table:paretotypes}. Moreover, the energy losses do slightly differ, as the conduction and switching losses are affected by the chip area difference of the main switches, as described in section \ref{sec:flo}. The maximum chip area data point of TNPC$_B$ (2L/3L) is equal to the location of the TNPC$_B$ (3L) design, as the required energy loss to chip area ratio is equal. It is important to note that the B6$^2$-Y inverter is not capable of the maximum load in Y-mode, as evidenced by the operating limits depicted in Fig. \ref{fig:PL_Pmoth}.  As illustrated in subfigure \textbf{(b)}, the utilization of optimized switching frequencies results in a slight reduction in energy losses for all topologies. The elliptical dashed lines serve as a reference for loss reduction between subplots \textbf{(a)} and \textbf{(b)}. The loss reduction aligns with the conclusions of previous studies \cite{velic2021efficiency} that have examined the overall efficiency of a drivetrain in relation to variations in switching frequency.

% %%%%%%%%%%%%%%%%%%%%%%%%%%%%%%%%%%% CONCLUSION %%%%%%%%%%%%%%%
\section{Conclusion}

The optimization of various inverter topologies has provided valuable insights into the interplay between additional cost due to increased chip area and efficiency. With a modest increase in semiconductor chip area, significant improvements in system efficiency have been demonstrated through optimal switching between different operating modes. During partial load operation, the study highlights the impact of motor harmonic losses and switching frequency on total system losses. Advanced control strategies and modulation techniques play a crucial role in mitigating these losses and optimizing power conversion across varying load conditions. The comparative analysis of B6, TNPC, and B6$^2$-Y reveals that each topology offers distinct advantages depending on application requirements and operational conditions. The B6 inverter is still unbeaten regarding minimum total chip area requirements, but leads to higher energy losses. Under partial load scenarios, the TNPC topology demonstrates superior performance in minimizing power losses through its 3-level operation. Conversely, the B6$^2$-Y topology in open-winding motor configurations, leverages the full DC-Link voltage to optimize efficiency without excessive increase of semiconductor chip area and easily allows the combination of SiC MOSFETs for partial load in Y-mode and Si IGBTs \& SiC MOSFETs under full load. The cost-benefit ratio of chip area and efficiency increase is highly dependent on the vehicle type, as well as desired driving range and must ultimately be considered by the manufacturer. However, the Pareto fronts in Fig. \ref{fig:pareto} and the operation boundary plots in Fig. \ref{fig:PL_opswitch} imply that a chip area increase of up to 50\,\% more than B6 is particularly sensible, as this is where the majority of the efficiency increase are seen. Above 50\,\% additional chip area, the benefits for a 300\,kW drive flatten out. A good compromise between max. efficiency and low cost could be around 25\,\% to 30\,\% additional chip area for the TNPC inverter design. Future research directions could focus on refining modulation and control strategies or considering novel inverter-motor topologies to push the limits of powertrain performance and efficiency even further.

% References
{\setstretch{1}\vspace{\baselineskip}
\bibliographystyle{IEEEtran}
\bibliography{00_main}
}

% \vspace{12pt}
% \color{red}
% IEEE conference templates contain guidance text for composing and formatting conference papers. Please ensure that all template text is removed from your conference paper prior to submission to the conference. Failure to remove the template text from your paper may result in your paper not being published.

\end{document}